# Discretization of Natanzon potentials

Artur Ishkhanyan[1] and Vladimir Krainov[2]

[1]Institute for Physical Research, NAS of Armenia, 0203 Ashtarak, Armenia
[2]Moscow Institute of Physics and Technology, 141700 Dolgoprudny, Russia

**Abstract.** We show that the Natanzon family of potentials is necessarily dropped into a restricted set of distinct potentials involving a fewer number of independent parameters if the potential term in the Schrödinger equation is proportional to an energy-independent parameter and if the potential shape is independent of both energy and that parameter. In the hypergeometric case only six such potentials exist, all five-parametric. Among these, only two (Eckart, Pöschl-Teller) are independent in the sense that each cannot be derived from the other by specifications of the involved parameters. Discussing the solvability of the Schrödinger equation in terms of the single-confluent Heun functions, we show that in this case there exist in total fifteen seven-parametric potentials, of which independent are nine. Six of the independent potentials present different generalizations of the hypergeometric or confluent hypergeometric ones, while three others do not possess hypergeometric sub-potentials. The result for the double- and bi-confluent Heun equations produces the three independent double- and five independent bi-confluent six-parametric Lamieux-Bose potentials, and the general five-parametric quartic oscillator potential for the tri-confluent Heun equation.

**1. Introduction.** – We consider the reduction of the stationary Schrödinger equation to a standard linear second-order ordinary differential equation having rational coefficients via transformation of the independent and dependent variables. If the coordinate transformation $z(x)$ is such that $dz/dx = \Pi_i (z-z_i)^{A_i}$ with constant $z_i$ and $A_i$, the potentials derived in this way can be referred to as the Manning potentials who discussed the reduction of the Schrödinger equation to a rather large class of target equations by a transformation of this form involving three factors: $z'(x) = z^{A_1}(1-z)^{A_2}(z-z_3)^{A_3}$ [1]. Examples of potentials derived by the Manning scheme include the harmonic oscillator, inverse square and Coulomb potentials discussed by Schrödinger [2], Manning [1], Eckart [3], Rosen-Morse [4], Woods-Saxon [5], Manning-Rosen [6], Hulthén [7], Pöschl-Teller [8] Scarf [9], Kratzer [10], Morse [11], Morse-Stückelberg [12], Manning [13], and many others explored in the past [13-16]. Recent examples include, e.g., the reduction of the Schrödinger equation to the five Heun equations or their certain extensions [17-20].

A further specification of the potentials concerns the cases when the coordinate transformation does not depend on the energy. The potentials obeying this restriction may be referred to as the Natanzon potentials [16], who explored the general conditions for the potentials to be solvable, under this supposition, in terms of the ordinary and confluent hypergeometric functions. For the hypergeometric equation, Natanzon constructed a parametrically given potential with the coordinate transformation defined through the equation

$$z'(x) = \frac{z(1-z)}{\sqrt{r_0 + r_1 z + r_2 z^2}}. \qquad (1)$$

The polynomial under the root can be written as $r_0$, $r_1(z-z_1)$, $r_2(z-z_1)(z-z_2)$, hence, this transformation can be viewed as a Manning transformation with at most four factors. Since the potential is given in parametric form, some of its properties are not immediately seen.

In the present paper we show that if the potential in a coordinate system is proportional to a parameter which is independent of energy $E$, that is, the potential term in the Schrödinger equation is presented as $mV(x) = \mu f(x)$ with $\mu \neq \mu(E)$, and if $f \neq f(\mu, E)$, some parameters involved in a Natanzon transformation necessarily adopt discrete values from a restricted permissible set.

This specification concerns the roots of the polynomial $r(z)$, and it comes from the singularities of the target equation to which the Schrödinger equation is reduced. More specifically, the assertion is that $r_{0,1,2}$ should be such that the roots of the polynomial $r(z)$ coincide with the singularities of the target equation. In particular, in the hypergeometric case only the roots $z_{1,2} = 0,1$ are permitted. Then, there are six possibilities: $r(z) \sim 1, z, 1-z, z^2, z(1-z)$, $(1-z)^2$. Because of the symmetry with respect to interchange $z \leftrightarrow 1-z$, the number of independent cases is reduced to four. Interestingly, it further turns out that there exist only two independent hypergeometric potentials which cannot be transformed into each other by specification of the remaining three involved parameters. As such independent potentials, one may choose the Eckart and Pöschl-Teller potentials derived, e.g., by $r = r_0$ and $r = r_1 z$, respectively. In the confluent hypergeometric case there are three independent potentials (Morse, oscillator, Kratzer), which are derived by choosing $r = r_0$, $r = r_1 z$ and $r = r_2 z^2$.

Mathematically, we first show that under the supposition that the parameter $\mu$ does not depend on energy and $f(x) \neq f(\mu, E)$, the parameters $r_i$ cannot depend on $\mu$. Note that for equations more general than the hypergeometric equations the polynomial $r(z)$ generally becomes of higher degree so that the number of involved parameters is increased (see below the example of the confluent Heun equation when the polynomial is of the fourth degree). It further follows that all the roots $z_i$ of polynomial $r(z)$ should necessarily coincide with the (finite) singular points of the target equation to which the Schrödinger equation is reduced. Then, the coordinate transformation becomes of the form $z'(x) = z^{k_1}(z-1)^{k_2}(z-z_3)^{k_3}... = \Pi_i(z-z_i)^{k_i}$ with $z_i$ being the singular points of the target equation. Finally, we see that the exponents $k_i$ should be integers or half-integers.

As an application, we consider the reduction of the Schrödinger equation to the four confluent Heun equations [21].

The result for the single-confluent Heun equation is that there exist fifteen possible choices for the coordinate transformation each leading to a seven-parametric potential. Because of the symmetry of the confluent Heun equation with respect to the interchange $z \leftrightarrow 1-z$, however, the number of independent potentials is effectively reduced to nine. Five of these potentials present different generalizations of either ordinary or confluent hypergeometric potentials, and one of the potentials has hypergeometric sub-cases of both types. Unlike the hypergeometric case, the confluent Heun potentials in general cannot be transformed into each other by specification of the involved parameters. Among the three potentials that do not possess hypergeometric sub-cases, a potential is explicitly written through a coordinate transformation given in terms of the Lambert $W$-function, which is an implicitly elementary function also known as the product logarithm [22].

Finally, we show that the Schrödinger equation is reduced to the double-confluent Heun equation for three independent potentials, five other independent potentials allow reduction to the bi-confluent Heun equation, and for the tri-confluent Heun equation there exists only one potential, the quartic oscillator. The double- and bi-confluent Heun potentials have been presented by Lamieux and Bose [17].

**2. The Natanzon potentials.** The one-dimensional stationary Schrödinger equation for a particle of mass $m$ and energy $E$ in a potential $V(x)$ is written as

$$\frac{d^2\psi}{dx^2} + \frac{2m}{\hbar^2}(E - V(x))\psi = 0. \quad (2)$$

Applying the transformation of the independent variable $z = z(x)$, this equation is rewritten for the argument $z$ as

$$\psi_{zz} + \frac{\rho_z}{\rho}\psi_z + \frac{2m}{\hbar^2}\frac{E-V(z)}{\rho^2}\psi = 0, \quad (3)$$

where (and hereafter) the lowercase Latin index denotes differentiation and $\rho = dz/dx$. Further transformation of the dependent variable $\psi = \varphi(z)u(z)$ reduces this equation to the following one for the new dependent variable:

$$u_{zz} + \left(\frac{2\varphi_z}{\varphi} + \frac{\rho_z}{\rho}\right)u_z + \left(\frac{\varphi_{zz}}{\varphi} + \frac{\rho_z}{\rho}\frac{\varphi_z}{\varphi} + \frac{2m}{\hbar^2}\frac{E-V}{\rho^2}\right)u = 0$$

(4)

We explore the cases when this equation becomes a target equation given as

$$u_{zz} + f(z)u_z + g(z)u = 0. \quad (5)$$

Thus, we demand

$$2\frac{\varphi_z}{\varphi} + \frac{\rho_z}{\rho} = f(z), \quad (6)$$

$$\frac{\varphi_{zz}}{\varphi} + \frac{\rho_z}{\rho}\frac{\varphi_z}{\varphi} + \frac{2m}{\hbar^2}\frac{E-V(z)}{\rho^2} = g(z). \quad (7)$$

From equation (6) we have

$$\varphi(z) = \rho^{-1/2} \exp\left(\int f(z)dz/2\right). \quad (8)$$

With this, equation (7) is rewritten as

$$g - \frac{f_z}{2} - \frac{f^2}{4} = -\frac{1}{2}\left(\frac{\rho_z}{\rho}\right)_z - \frac{1}{4}\left(\frac{\rho_z}{\rho}\right)^2 + \frac{2m}{\hbar^2}\frac{E-V(z)}{\rho^2}. \quad (9)$$

On the left-hand side of this equation we recognize the *invariant* of (5) if it is transformed to its *normal* form:

$$I(z) = g - \frac{f_z}{2} - \frac{f^2}{4}, \quad (10)$$

and the first two terms on the right-hand side are identified as

$$-\frac{1}{2}\left(\frac{\rho_z}{\rho}\right)_z - \frac{1}{4}\left(\frac{\rho_z}{\rho}\right)^2 = -\frac{\{z,x\}}{2\rho^2}, \quad (11)$$

where $\{z, x\}$ is the Schwartzian derivative.

Thus, equation (9) is rewritten as

$$z'(x)^2 I(z) + \frac{1}{2}\{z,x\} = \frac{2m}{\hbar^2}(E - V(x)) \quad (12)$$

(the prime denotes differentiation with respect to $x$). This is the equation considered by Bose [14] and Natanzon [16]. Sometimes it is convenient to use equation (9) which employs the *canonical* forms of equations obeyed by the standard special functions. The latter forms are written in an intuitive way that reveals the singularity structure of the equations, and it is easier to apply, without intermediate transformations, the standard mathematical knowledge. It is, however, understood that the two techniques are completely equivalent.

The Natanzon approach for constructing solvable potentials rests on the supposition that the coordinate transformation is energy-independent. In general, this is not a necessary condition; many non-Natanzon potentials solvable in terms of standard special functions are known. Still, the approach provides a major set of known exactly solvable potentials, including the classical cases. Within the supposition that $z(x)$ is $E$-independent, it is immediately seen that the energy and potential terms in (12) should independently match the left-hand side term involving the invariant $I(z)$. For the Gauss hypergeometric equation:

$$u_{zz} + \left(\frac{\gamma}{z} + \frac{\delta}{z-1}\right)u_z + \frac{\alpha\beta z}{z(z-1)}u = 0, \quad (13)$$

the invariant is a second-degree polynomial in $z$ divided by $z^2(1-z)^2$. Hence, one straightforwardly arrives at the transformation (1). The Natanzon potential is then given as

$$2mV(x)/\hbar^2 = \frac{v_0 + v_1 z + v_2 z^2}{r_0 + r_1 z + r_2 z^2} - \frac{\{z,x\}}{2}. \quad (14)$$

With arbitrary $r_{0,1,2}$, $v_{0,1,2}$ and the integration constant of equation (1), this is a seven-parametric family of potentials. However, the classical hypergeometric potentials are achieved by such specifications of the parameters $r_{0,1,2}$ that $r(z) \sim 1$, $z$, $1-z$, $z^2$, $z(1-z)$ or $(1-z)^2$ [14].

The confluent hypergeometric potentials are derived in the same manner starting from the *scaled* Kummer confluent hypergeometric equation of the form

$$u_{zz} + \left(\frac{\gamma}{z} + \varepsilon\right) u_z + \frac{\alpha}{z^2} u = 0. \quad (15)$$

The result just slightly differs from that for the previous case: $z'(x) = z/\sqrt{r(z)}$ and $V(x)$ is given by equation (14). Hence, this is also a seven-parametric potential, however, the classical confluent hypergeometric potentials are derived by rather severe specifications: $r(z) \sim 1$, $z$ or $z^2$ [14].

Because of this specification of the parameters $r_{0,1,2}$ the potentials become five-parametric. Since three of the parameters stand for the coordinate origin, the space scale and the origin of the energy, there remain only two parameters for characterization of the potentials. In the next section we consider a situation that necessarily leads to such discretization of the Natanzon potentials.

**3. Discretization of Natanzon potentials.** We consider the particular case when the potential term in the Schrödinger equation is proportional to a parameter which does not depend on energy and suppose that the potential *shape* is independent of both energy and that parameter. Thus, we suppose

$$2mV(x)/\hbar^2 = \mu f(x) \quad (16)$$

with $\mu \neq \mu(E)$ and $f \neq f(\mu, E)$. As it is readily seen, an example of such a situation is the case of pure electromagnetic interactions when one can just put $\mu = m$.

Consider the hypergeometric case. To show the discretization, one needs to prove that the roots of the polynomial $r(z)$ coincide with the singularities of the hypergeometric equation $z = 0, 1$. As a first step we ask if it is possible for the potential given by equation (14) to be of the form (16) if the transformation defined by equation (1) is $\mu$-dependent. The answer is no. To show this, we consider the behavior of the $\mu$-independent functions $f(x)$ and $F(x) = \ln f'(x)$. The latter function provides the following identity

$$F'(x) = \frac{V''(x)}{V'(x)} = \frac{V_{zz}(z)}{V_z(z)} \rho(z) + \rho_z(z), \quad (17)$$

which is rather useful for what follows. Indeed, taking the limits $z \to 0, 1$, it is shown that $r(0)$ and $r(1)$ are $\mu$-independent.

Let $r_0 \neq 0$. Then, since $r(0) = r_0$, we reveal that $r_0$ does not depend on $\mu$. According to equation (1), we have the following power-series expansion in the vicinity of $z = 0$:

$$\frac{dx}{dz} = \frac{1}{z'(x)} = \frac{\sqrt{r_0}}{z} + a_0 + a_1 z + a_2 z^2 + \dots \quad (18)$$

Retaining only the first term on the right-hand side of this equation, we get that the leading asymptote is

$$z\big|_{x \to -\infty} \sim \exp\left(\frac{x - x_0}{\sqrt{r_0}}\right) = c_0 \exp\left(x/\sqrt{r_0}\right), \quad (19)$$

which allows establishing the following useful limit for any $n$:

$$\lim_{x \to -\infty} e^{-nx/\sqrt{r_0}} z^n(x) = c_0^n, \quad c_0 = \exp\left(-x_0/\sqrt{r_0}\right). (20)$$

Using equation (14), the functions $f(x)$ and $F(x)$ allow the expansions

$$f(x) = \frac{2m}{\mu \hbar^2}(V_0 + V_1 z + V_2 z^2 + \dots)$$
$$= f_0 + f_1 z + f_2 z^2 + \dots \quad (21)$$

$$F(x) = F_0 + F_1 z + F_2 z^2 + \dots \quad (22)$$

Considering the partial sums $P_n(x) = \Sigma_{k=0}^{n} f_k z^k$, with $P_{-1} \equiv 0$, and taking the successive limits

$$\lim_{x \to -\infty} e^{-nx/\sqrt{r_0}} \left( f(x) - P_{n-1}(x) \right) = f_n c_0^n \quad (23)$$

for $n = 0, 1, 2, \ldots$, we conclude, since $f(x)$ is $\mu$-independent, that $f_n c_0^n$ do not depend on $\mu$. Similarly, all $F_n c_0^n$ are $\mu$-independent. The coefficients $F_n$ can be expressed in terms of $f_n$. The first two coefficients read

$$F_0 = \frac{2 f_2 \rho(0)}{f_1} + \rho_z(0) = \frac{1}{\sqrt{r_0}},$$

$$F_1 = \frac{2 f_2}{\sqrt{r_0} f_1} - \frac{2 r_0 + r_1}{r_0^{3/2}}. \quad (24)$$

These are informative equations. The first equation confirms that $r_0$ is $\mu$-independent. Further, the second equation and the one for the next coefficient, produce the equations:

$$\frac{d(c_0 F_1)}{d\mu} = 0 = \frac{d}{d\mu} \left( c_0 (2 r_0 + r_1) \right), \quad (25)$$

$$\frac{d(c_0^2 F_2)}{d\mu} = 0 = \frac{d}{d\mu} \left( c_0^2 \left( A(r_{0,1}) + B(r_{0,1,2}) \right) \right). (26)$$

from which, together with the information that $r_0 + r_1 + r_2 = r(1)$ is $\mu$-independent, it is readily established that $r_{0,1,2}$ and $c_0$ are $\mu$-independent. Thus, if $r_0 \neq 0$, the coordinate transformation is $\mu$-independent. The case $r_0 = 0$ is almost trivial because the $\mu$-independence of $r(0)$ and $r(1)$ immediately shows that $r_{1,2}$ are $\mu$-independent so that equation (25) suffices to see that $c_0$, and thus the coordinate transformation $z(x)$, is $\mu$-independent. A last remark is that though we used information from the singular point $z = 1$, in fact, it is sufficient to consider only the behavior in the vicinity of $z = 0$. Indeed, instead of the $\mu$-independence of $r(1)$ one can apply the equation $d(c_0^3 F_3)/d\mu = 0$.

The above derivations are exact, and the same approach is applied to any differential equation having a singularity located at a finite point of complex $z$-plane. Hence, the $\mu$-independence for $z(x)$ is a general result for the Natanzon potentials if the potential term is proportional to an energy-independent parameter $\mu$, and if the potential shape is both $E$- and $\mu$-independent. The result holds with proviso that the Schrödinger equation is reduced to an equation, which has a singularity located at a finite point, and that the variation range of $z$ includes the vicinity of this point.

Thus, the transformation $z(x)$ is $\mu$-independent. The last step is then straightforward. Indeed, taking the limits $E \to 0$ and $\mu \to 0$ in equation (9), we have

$$\frac{1}{2}\left(\frac{\rho_z}{\rho}\right)_z + \frac{1}{4}\left(\frac{\rho_z}{\rho}\right)^2 = -\left( g - \frac{f_z}{2} - \frac{f^2}{4} \right)_{E,\mu \to 0} .(27)$$

It is then immediately seen, that the function $\rho_z / \rho$ cannot have poles other than the singularities of the invariant $I(z)$. Consequently, the roots of the polynomial $r(z)$ should coincide with the singularities of equation (5).

**4. Confluent Heun potentials.** – As a representative example, we first apply the above theorem to the single-confluent Heun equation which is a second-order differential equation having regular singularities at $z = 0, 1$ and an irregular singularity of rank 1 at $z = \infty$. In its canonical form, the equation is written as [21]

$$u_{zz} + \left( \frac{\gamma}{z} + \frac{\delta}{z-1} + \varepsilon \right) u_z + \frac{\alpha z - q}{z(z-1)} u = 0. \quad (28)$$

Among the five Heun equations, this equation is of special interest because it directly incorporates the hypergeometric and (scaled) confluent hypergeometric equations widely applied in quantum mechanics in the past. The ordinary hypergeometric equation is achieved by putting $\varepsilon = \alpha = 0$, while the confluent one is the case if $\delta = 0$ and $q = \alpha$.

Since the finite singularities are $z = 0, 1$, the Natanzon family of the confluent Heun potentials is constructed by the coordinate transformation $z(x)$ of the form

$$z'(x) = \rho = z^{m_1}(z-1)^{m_2}/\sigma \qquad (29)$$

with integer or half-integer $m_{1,2}$ and arbitrary constant $\sigma$. Substituting this into equation (9) and examining the energy term, we see that $z^2(z-1)^2/\rho^2$ is a polynomial in $z$ of at most fourth degree. This imposes the inequalities $1 \geq m_{1,2}$, $m_1 + m_2 \geq 0$, which lead to 15 possible sets of $m_{1,2}$ shown in Fig. 1 by points in 2D space $(m_1, m_2)$. The cases possessing hypergeometric sub-potentials are marked by squares, and those having confluent hypergeometric sub-potentials are marked by triangles. There exist two cases when the potential possesses hypergeometric sub-potentials of both types. These cases are marked by rhombs. Because the confluent Heun equation preserves its form when interchanging $z \leftrightarrow 1-z$, the number of independent potentials is only nine. These nine cases are indicated in Fig. 1 by filled shapes.

Finally, examining the potential term in equation (9), we see that $z^2(z-1)^2 V(z)/\rho^2$ is also a polynomial in $z$ of at most fourth degree. Thus, we get

$$V(z) = z^{2m_1-2}(z-1)^{2m_2-2}(v_0 + v_1 z + v_2 z^2 + v_3 z^3 + v_4 z^4)$$
$$. \qquad (30)$$

With $m_{1,2}$ from Fig. 1, this equation defines 15 seven-parametric potentials, the independent 9 of which can be written in the form presented in Table 1. Six of these potentials, those possessing hypergeometric sub-potentials, were presented by Lamieux and Bose [15] (the ten potentials listed in Table II, p. 265 of [15] are particular cases of the mentioned six, e.g., the last three rows of Table II are specifications of the potential with $m_{1,2} = (1,1)$). Some non-hypergeometric representatives of these six potentials have been discussed by many authors on several occasions, e.g., in connection with the two-centre Coulomb problem or Teukolsky equation [23]. It seems that the three remaining potentials have not been treated before. One of these potentials, that with $m_{1,2} = (1,-1)$, is explicitly written in terms of the Lambert function, which is an implicitly elementary function that resolves the equation $W \exp(W) = x$ [22].

The solution of the Schrödinger equation for the presented potentials is written in terms of the confluent Heun function as

$$\psi = e^{\alpha_0 z} z^{\alpha_1}(z-1)^{\alpha_2} H_C(\gamma, \delta, \varepsilon; \alpha, q; z). \qquad (31)$$

Substituting this into equations (6) and (7) and collecting the coefficients at powers of $z$, we get eight equations which are linear for the five parameters of the confluent Heun function and are quadratic for the three parameters $\alpha_{0,1,2}$ of the pre-factor. Resolving these equations is strightforward.

Thus, for any row of Table 1, we have a potential given parametrically as a pair of functions $x(z), V(z)$. The transformation $x(z)$ in general can be written in terms of the incomplete Beta-functions. However, since the parameters $m_{1,2}$ are integers or half-integers, the Beta-functions are always reduced to elementary functions. These functions are presented in the third column of Table 1.

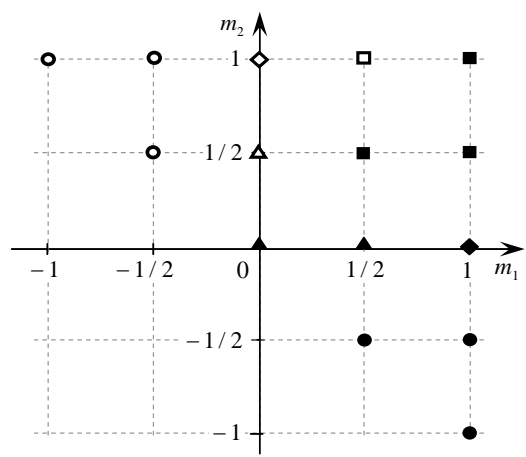

Fig. 1. Fifteen possible pairs $(m_1, m_2)$.

| $m_1, m_2$ | Potential $V(z)$ | $(x-x_0)/\sigma$ | |
|---|---|---|---|
| 0, 0 | $V_0 + \dfrac{V_1}{z} + \dfrac{V_2}{z^2} + \dfrac{V_3}{z-1} + \dfrac{V_4}{(z-1)^2}$ | $z$ | $_1F_1$ [1,10] |
| 1/2, −1/2 | $V_0 + \dfrac{V_1}{z} + \dfrac{V_2}{z-1} + \dfrac{V_3}{(z-1)^2} + \dfrac{V_4}{(z-1)^3}$ | $\sqrt{z(z-1)} - \sinh^{-1}(\sqrt{z-1})$ | |
| 1/2, 0 | $V_0 + V_1 z + \dfrac{V_2}{z} + \dfrac{V_3}{z-1} + \dfrac{V_4}{(z-1)^2}$ | $2\sqrt{z}$ | $_1F_1$ [1] |
| 1/2, 1/2 | $V_0 + V_1 z + V_2 z^2 + \dfrac{V_3}{z} + \dfrac{V_4}{z-1}$ | $2\sinh^{-1}(\sqrt{z-1})$ | $_2F_1$ [8] |
| 1, −1 | $V_0 + \dfrac{V_1}{z-1} + \dfrac{V_2}{(z-1)^2} + \dfrac{V_3}{(z-1)^3} + \dfrac{V_4}{(z-1)^4}$ | $z - \log(z)$ | Lambert $W$ |
| 1, −1/2 | $V_0 + V_1 z + \dfrac{V_2}{z-1} + \dfrac{V_3}{(z-1)^2} + \dfrac{V_4}{(z-1)^3}$ | $2\sqrt{z-1} - 2\tan^{-1}(\sqrt{z-1})$ | |
| 1, 0 | $V_0 + V_1 z + V_2 z^2 + \dfrac{V_3}{z-1} + \dfrac{V_4}{(z-1)^2}$ | $\log(z)$ | $_1F_1$ [11]  $_2F_1$ [3] |
| 1, 1/2 | $V_0 + V_1 z + V_2 z^2 + V_3 z^3 + \dfrac{V_4}{z-1}$ | $2\tan^{-1}(\sqrt{z-1})$ | $_2F_1$ [8] |
| 1, 1 | $V_0 + V_1 z + V_2 z^2 + V_3 z^3 + V_4 z^4$ | $2\tanh^{-1}(1-2z)$ | $_2F_1$ [3] |

Table 1. Nine independent seven-parametric confluent Heun potentials.

In the six cases possessing hypergeometric sub-potentials the inverse transformation $z(x)$ is written in terms of elementary functions. Among these, specific is the case $m_{1,2} = (1, 0)$ for which $z = \exp((x - x_0)/\sigma)$ and the confluent Heun potential presents the sum of the Morse confluent hypergeometric [11] and the Eckart ordinary hypergeometric [3] potentials:

$$V = V_0 + V_1 e^{\frac{x-x_0}{\sigma}} + V_2 e^{2\frac{x-x_0}{\sigma}} + \frac{V_3}{e^{\frac{x-x_0}{\sigma}} - 1} + \frac{V_4}{(e^{\frac{x-x_0}{\sigma}} - 1)^2}. \quad (32)$$

Among the three cases of the lower right quadrant in Fig.1 that do not possess hypergeometric sub-potentials, we note the potential with $m_{1,2} = (1, -1)$:

$$V(z) = \sum_{n=0}^{4} \frac{V_n}{(z-1)^n}, \quad (33)$$

for which the transformation $z(x)$ is written in terms of an *implicitly elementary* function, the Lambert $W$-function [22]:

$$z = -W\left(-e^{-(x-x_0)/\sigma}\right). \quad (34)$$

For a positive $\sigma$ and $x_0 = -\sigma$ this is a potential defined on the positive half-axis $x > 0$. The potential has a singularity at the origin and tends to $V_\infty = \Sigma_{n=0}^{n=4}(-1)^n V_n$ at infinity. In the vicinity of the origin applies the expansion

$$V\big|_{x \to 0} = \frac{d_{-4}}{x^2} + \frac{d_{-3}}{x^{3/2}} + \frac{d_{-2}}{x} + \frac{d_{-1}}{x^{1/2}} + d_0 + \ldots \quad (35)$$

Because equation (33) involves five independent parameters, the first five parameters in this expansion may adopt arbitrary values. The behavior of the potential at infinity is

$$V\big|_{x \to +\infty} = V_\infty - A e^{-(x+\sigma)/\sigma}, \quad (36)$$

where $A = V_1 - 2V_2 + 3V_3 - 4V_4$.

**5. Double-, bi- and tri-confluent Heun potentials.** – To treat the cases of the double-confluent (DHE), bi-confluent (BHE) and tri-confluent (THE) Heun equations, it is convenient to adopt the following canonical forms for these equations which differ from those applied in the standard references [1,2] in that they provide unified *five-parametric* representation for all four confluent Heun equations:

DHE: $u_{zz} + \left(\dfrac{\gamma}{z^2} + \dfrac{\delta}{z} + \varepsilon\right) u_z + \dfrac{\alpha z - q}{z^2} u = 0$ (36)

BHE: $u_{zz} + \left(\dfrac{\gamma}{z} + \delta + \varepsilon z\right) u_z + \dfrac{\alpha z - q}{z} u = 0$ (37)

THE: $u_{zz} + \left(\gamma + \delta z + \varepsilon z^2\right) u_z + (\alpha z - q) u = 0$ (38)

(compare with equation (28)). Though the number of irreducible parameters in these three equations is less than five (four for DHE and BHE and three for THE), however, these forms provide a convenient unified description of results which allows one to easily follow the links between different equations.

The reduction of the Schrödinger equation to equations (36)-(38) is straightforward. Indeed, since the double- and bi-confluent equations possess only one finite singularity, conventionally, located at $z = 0$, the permissible coordinate transformation for these equations is given through the equation

$$z'(x) = \rho = z^{m_1} / \sigma \qquad (39)$$

with integer or half-integer $m$. Hence,

$z \sim x^{1/(1-m_1)}$ if $m_1 \neq 1$ and $z \sim e^{x/\sigma}$ if $m_1 = 1$. (40)

The invariants of the double-confluent and bi-confluent equations are fourth-degree polynomials divided by $z^4$ and $z^2$, respectively. Accordingly, for the exponent $m_1$ we have

$0 \leq 4 - 2m_1 \leq 4$ and $0 \leq 2 - 2m_1 \leq 4$ (41)

for DHE and BCE, respectively. Hence, $m_1 = 0, 1/2, 1, 3/2, 2$ for DHE, and for BCE we have $m_1 = -1, -1/2, 0, 1/2, 1$.

The corresponding potentials are constructed by applying the equation

$$V = z^{2m_1-d}(v_0 + v_1 z + v_2 z^2 + v_3 z^3 + v_4 z^4), \quad (42)$$

where $d = 4$ for the double-confluent case and $d = 2$ for the bi-confluent equation. Using equations (40), the potentials are written as explicit functions of $x$. It turns out that for the double-confluent case independent are only the first three potentials with $m_1 = 0, 1/2, 1$.

In general, all the double- and bi-confluent Heun potentials involve six independent parameters. All these potentials have been previously presented by Lamieux and Bose [15]. For the convenience of the reader, we reproduce the potentials in Table 2 where we have omitted the parameters $x_0, \sigma$ (everywhere one should replace $x \to (x - x_0)/\sigma$).

| $m_1$ | Double-confluent Heun potentials | $m_1$ | Bi-confluent Heun potentials |
|---|---|---|---|
| | | $-1$ | $V(x) = \dfrac{V_0}{x^2} + \dfrac{V_1}{x^{3/2}} + \dfrac{V_2}{x} + \dfrac{V_3}{x^{1/2}} + V_4$ |
| | | $-1/2$ | $V(x) = \dfrac{V_0}{x^2} + \dfrac{V_1}{x^{4/3}} + \dfrac{V_2}{x^{2/3}} + V_3 + V_4 x^{2/3}$ |
| $0$ | $V_0 + \dfrac{V_1}{x} + \dfrac{V_2}{x^2} + \dfrac{V_3}{x^3} + \dfrac{V_4}{x^4}$ | $0$ | $V(x) = \dfrac{V_0}{x^2} + \dfrac{V_1}{x} + V_2 + V_3 x + V_4 x^2$ |
| $1/2$ | $V_0 x^2 + V_1 + \dfrac{V_2}{x^2} + \dfrac{V_3}{x^4} + \dfrac{V_4}{x^6}$ | $1/2$ | $V(x) = \dfrac{V_0}{x^2} + V_1 + V_2 x^2 + V_3 x^4 + V_4 x^6$ |
| $1$ | $V_0 e^{-2x} + V_1 e^{-x} + V_2 + V_3 e^{x} + V_4 e^{2x}$ | $1$ | $V(x) = V_0 + V_1 e^{x} + V_2 e^{2x} + V_3 e^{3x} + V_4 e^{4x}$ |

Table 2. Three double-confluent and five bi-confluent potentials.

The case of the tri-confluent Heun equation is different because this equation does not possess a finite singularity. It is then understood that in this case the polynomial $r(z)$ should not have roots at all. Then, the only possibility is $r(z) = r_0 = \text{const}$ so that the only possible coordinate transformation is $z = (x - x_0)/\sigma$. The corresponding tri-confluent Heun potential is the general quartic oscillator:

$$V(x) = V_0 + V_1 x + V_2 x^2 + V_3 x^3 + V_4 x^4, \quad (43)$$

where one may replace $x \to (x - x_0)/\sigma$. However, as it is immediately seen, this replacement does not change the general form of (43), hence, this is a five-parametric potential.

**6. Discussion.** – The solution of the Schrödinger equation in terms of special mathematical functions has a long history [1-21]. The general algebraic form of the independent variable transformation was discussed by Manning [1] who, however, did not consider the details due to the pre-factor.

Discussing the reducibility to the hypergeometric equations, Natanzon presented a general analysis, with the pre-factor, for the case when the coordinate transformation is energy-independent [16]. His result is a potential involving seven parameters, all supposed to be continuous.

The main result of the present paper is that if the potential is proportional to a parameter $\mu$, which does not depend on the energy, and if the potential shape is independent of both energy and this parameter, the Natanzon family of potentials is dropped into a finite set of separate potentials involving fewer continuous parameters. We first show that the coordinate transformation $z(x)$ applied to construct the Natanzon potentials via reduction of the Schrödinger equation to an equation, which has a finite singularity, should be $\mu$-independent. Further, we show that then the coordinate transformation should necessarily be of the Manning form with factors involving only the singularities of the target equation. We note that our result is general and applies to any equation possessing a singularity located at a finite point of complex $z$-plane. The result implies that the variation range of $z$ includes the vicinity of the finite singular point of the considered equation.

The discretization of the Natanzon potentials has been noticed on several occasions. For example, the authors of [24], have shown that the discretization is necessarily the case if the potential term is proportional to an independent parameter $\mu$; however, their discussion rests on the *presupposition* that the independent variable transformation is $\mu$-independent. The complementary observation we report here is that the $\mu$-independence of the coordinate transformation is necessarily the case for any target equation possessing a singular point if the variable $z$ is allowed to vary in a region that includes the vicinity of this singularity.

As a representative example, we have discussed the confluent Heun equation which presents a natural generalization of the two hypergeometric equations. Like the quantum two-state problem [25], there exist fifteen seven-parametric potentials for which the Schrödinger equation can be reduced to the confluent Heun equation. However, only nine potentials are independent. Six of these independent cases suggest different generalizations of the hypergeometric sub-potentials, while the other three do not allow hypergeometric reductions. Unlike the case of the ordinary hypergeometric sub-potentials, no confluent Heun potential can be transformed into another one by means of specification of the involved parameters. Among the potentials that do not posses hypergeometric sub-cases, a potential has an explicit representation through a coordinate transformation written in terms of the Lambert $W$-function, which is an implicitly elementary function known also as the product logarithm.

Finally, we have shown that the $\mu$-independent Natanzon-type potentials reducible to the double-confluent, bi-confluent and tri-confluent Heun equations are the three six-parametric double- and five six-parametric bi-confluent Lamieux-Bose potentials [17] and one five-parametric tri-confluent Heun potential (the quartic oscillator).

**Acknowledgments**

This work has been supported by the Armenian State Committee of Science (Grant No. 13RB-052) and the Ministry of Education and Science of the Russian Federation (state assignment No. 3.679.2014/K).